\documentclass{edp-conf}
\usepackage{graphicx}
\begin{document}

\TitreGlobal{SF2A 2005}

%%-----------------------------
%%      the top matter
%%-----------------------------
\title{CO(2-1) large scale mapping of the Perseus cluster core with HERA}
\author{P. Salome}\address{IRAM , 300 rue de la piscine 38400
St Martin d'Heres.}  \author{F. Combes}
\address{LERMA, Observatoire de Paris, 61 Av. de l'Observatoire 75014
Paris.}
\runningtitle{CO filaments around NGC 1275}
\setcounter{page}{237}
% Keep this line, even if the page will be settled afterwards..
\index{Author1, A.}
\index{Author2, B.}
\index{Author3, C.}
% Repeat the authors here, this will help to make the final index

\maketitle
\begin{abstract}
Cold molecular gas has recently been found is several cooling flow
clusters cores with single dish telescopes. High spatial resolution
imaging of some of these clusters then revealed the peculiar
morphology and dynamics of the CO emission lines, pointing out a
perturbed very cold component in the cluster centers. We report
here the observations of NGC 1275, in the Perseus cluster of
galaxies. This object is the strongest cooling flow emitter in the
millimeter. The 9 dual polarization pixels of the HERA focal plane array, installed on
the 30m telescope, enabled to image the large scale emission of the
cold molecular gas which is found to follow the very peculiar Halpha
filamentary structure around the central galaxy. We discuss here this
association and the non-rotating dynamics of the cold gas that argue
for a cooling flow origin of the molecular component.
\end{abstract}
%
%%-----------------------------
%%      your text
%%-----------------------------
\section{Mapping the filamentary structure}
\begin{figure}[h]
   \centering
   \includegraphics[width=10cm]{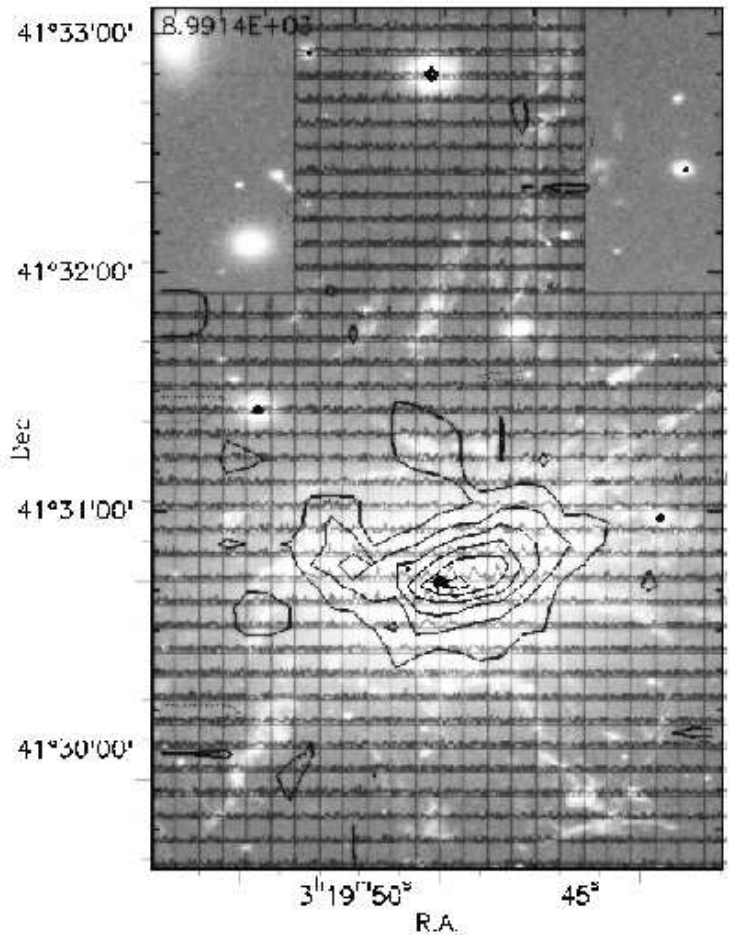}
      \caption{Halpha image from Conselice et al (2001), 
      with CO(2-1) contours superimposed. 
      Contours are linear, form 10 to 100\% of the maximum
      emission of 8.3 K km/s, in T$_{\rm A}^*$ scale. The boxes 
      show the spectra obtained with a velocity scale from -600 to 600 km/s, 
      and a temperature scale in T$_{\rm A}^*$ from -10 to 40 mK.}
       \label{figure1}
   \end{figure}

These observations were made from 1st to 3rd January 2005 at the IRAM-30m telescope.
We used the HERA (HEterodyne Receiver Array), a focal array of 18 SIS receivers,
9 for each polarisation, tuned at the CO(2-1) line, for NGC 1275 (226.56 GHz).
The sampling was 6 arsec (full sampling). We built four such maps, covering
the central 138x138$''$, and also a 5th one covering 66x66$''$ towards the north. 
Figure \ref{figure1} shows in contours the CO(2-1) emission, overlaid on a Halpha 
image of the filamentary structure pointed out by Conselice et al. (2001).

\section{Origin of the molecular gas}

The CO contours appear to surround a northern radio lobe cavity, also traced
in the X-ray gas.  The hot gas, compressed
towards the rims, cools there more efficiently, which
could explain the presence of CO gas. The kinematics deduced from the 
CO spectra is not regular, and it is not possible to follow a 
rotational pattern. The mass of cold gas found here 
(4 10$^{10}$ M$_\odot$) is quite a large amount for a single galaxy. 
So, it is likely that we see here, for the first time, a filamentary cold 
gas coming from a cooling flow as predicted by X-ray observations.

%%-----------------------------
%%      your bibliography
%%-----------------------------
%In preparing the reference list please adhere to the following format.
% Attention should be paid to the order of the items in each reference
% and to the punctuation used. Please see Sect. 4 in the User's Guide
% that comes with the new macro package.

%Bohr, N., Einstein, A., & Fermi, E. 1992, MNRAS, 301, 257 (BEF)
% Curie, M., & Curie, P. 1991, A&A, 248, 612
% de Gaulle, C. 1996, Solar Phys. (Oxford: Oxford Univ. Press)
% Heisenberg, W., & West, C. N. 1993, Australian J. Phys., 537, 36  (Paper III)
% Laurel, S., & Hardy, O. 1994, Active Galactic Nuclei, in The Evolution
% and Distribution of Galaxies, ed. W. Churchill, F. D. Roosevelt, & J.
% Stalin (New York: Wiley), 210


\begin{thebibliography}{}
\bibitem{Cons01} Conselice at al. 2001, AJ, 122, 2281   
\bibitem{Salo03} Salome, P. $\&$ Combes, F. 2003, A$\&$A, 412, 657   
\end{thebibliography}
\end{document}